%% file: main.tex
\documentclass{PoS}

\graphicspath{{./figs/}}
\usepackage{graphicx}
\usepackage[hyphens]{url}
\usepackage{cite}
\usepackage{array}
\usepackage{amsmath}
\usepackage{multirow}
\usepackage{caption}
\usepackage{subcaption}
\usepackage{mdframed}
\usepackage{color}
\usepackage[usenames,dvipsnames]{xcolor}
\usepackage{listings}
\title{Performance of Kepler GTX Titan GPUs and Xeon Phi System}

\ShortTitle{Performance of Kepler GTX Titan GPUs and Xeon Phi System}

\author{\speaker{Hwancheol Jeong}, Weonjong Lee, and Jeonghwan Pak \\
  Lattice Gauge Theory Research Center, CTP, and FPRD, \\
  Department of Physics and Astronomy, \\
  Seoul National University, Seoul, 151-747, South Korea \\
  E-mail: \email{wlee@snu.ac.kr}
}

\author{Kwang-jong Choi, Sang-Hyun Park, and Jun-sik Yoo \\
  Department of Physics and Astronomy, \\
  Seoul National University, Seoul, 151-747, South Korea
}

\author{Joo Hwan Kim, Joungjin Lee, and Young Woo Lee \\
  Seoul Science High School, Seoul, 110-530, South Korea
}

\input{abstract.tex}

\FullConference{31st International Symposium on Lattice Field Theory -
  LATTICE 2013\\ July 29 - August 3, 2013\\ Mainz, Germany}

\begin{document}


\input{intro.tex}


\input{titan.tex}


\input{gpuDir.tex}


\input{phi.tex}


\input{conclusion.tex}


\input{acknowledge.tex}


\bibliographystyle{JHEP}
\bibliography{ref}


\end{document}

%% file: abstract.tex
\abstract{
NVIDIA's new architecture, Kepler improves GPU's performance
significantly with the new streaming multiprocessor SMX.
Along with the performance, NVIDIA has also introduced many new
technologies such as direct parallelism, hyper-Q and GPU Direct with
RDMA.
Apart from other usual GPUs, NVIDIA also released another Kepler
`GeForce' GPU named GTX Titan.
GeForce GTX Titan is not only good for gaming but also good for
high performance computing with CUDA.
Nevertheless, it is remarkably cheaper than Kepler Tesla GPUs.
We investigate the performance of GTX Titan and find out how to
optimize a CUDA code appropriately for it.
Meanwhile, Intel has launched its new many integrated core (MIC) system,
Xeon Phi.
A Xeon Phi coprocessor could provide similar performance with NVIDIA
Kepler GPUs theoretically but, in reality, it turns out that its
performance is significantly inferior to GTX Titan.
}

%% file: intro.tex
\section{Introduction}

On 2013, NVIDIA launched a new Kepler GPU, GTX Titan, named after the
fastest supercomputer, a GPU cluster of NVIDIA Tesla K20X at Oak Ridge
National Laboratory \cite{web:Titan}.
GeForce GPUs are designed for gaming. However, GTX Titan is good for
parallel computing with CUDA, too.
From the standpoint of computing, GTX Titan is as great as Tesla K20X.
Nevertheless, the price of the former is about 3 times cheaper than of
the latter.

Meanwhile, in 2012, Intel announced the Xeon Phi system with Intel
many integrated core architecture (MIC) \cite{web:Phi}.
A Xeon Phi coprocessor integrates many CPU cores on a PCI express card
like GPU, so that it could, in principle, provide similar theoretical
performance with GTX Titan.
The merit is that most of usual C codes which runs on CPUs can run on
Xeon Phi system without much modification, because it is a CPU-based
platform.
However, it turns out that its performance is so low that it is very
hard to obtain the high performance from Xeon Phi.
%

%% file: titan.tex
\section{GTX Titan \& Kepler Architecture}


\begin{table}[tbhp]
  \centering
  \begin{tabular}{| >{\centering\arraybackslash}m{3.2cm} |
      >{\centering\arraybackslash}m{1.6cm} |
      >{\centering\arraybackslash}m{1.6cm} |
      >{\centering\arraybackslash}m{1.6cm} |
      >{\centering\arraybackslash}m{1.6cm} |
      >{\centering\arraybackslash}m{1.6cm} |}
    \hline
    Architecture & Fermi & Kepler (GK104) & \multicolumn{3}{c|} {Kepler
      (GK110)} \\
    \hline
    \multirow{2}{*}{GPU Device} & GTX & GTX & Tesla & GTX & GTX \\
    & 580 & 680 & K20X & TITAN & 780 \\
    \hline
    \# of CUDA Cores & 512 & 1536 & \color{blue} 2688 & \color{blue} 2688 &
      2304 \\
    \hline
    Core Clock {\scriptsize (MHz)} & 772 & 1006 & 732 & 837 & 863 \\
    \hline
    SP GFLOPS & 1581 & 3090 & 3950 & 4500 & 3977 \\
    \hline
    DP GFLOPS & 197 & 128 & \color{blue} 1312 & \color{blue} 1300 & 166 \\
    \hline
    Memory Size {\scriptsize (GB)} & 1.5 & 2 & \color{blue} 6 &
    \color{blue} 6.1 & 3 \\ 
    \hline
    Memory Bandwidth & \multirow{2}{*}{192.4} & \multirow{2}{*}{192.26} &
    \multirow{2}{*}{250} & \multirow{2}{*}{288.4} & \multirow{2}{*}{288.4} \\ 
    {\scriptsize (GB/sec)} & & & & & \\
    \hline
    L1 cache + & \multirow{2}{*}{64} & \multirow{2}{*}{64} &
    \multirow{2}{*}{64} & \multirow{2}{*}{64} & \multirow{2}{*}{64} \\ 
    shared memory {\scriptsize (KB)} & & & & & \\
    \hline
    read-only & \multirow{2}{*}{0} & \multirow{2}{*}{0} & \color{blue}
    \multirow{2}{*}{48} & \color{blue} \multirow{2}{*}{48} & \color{blue}
    \multirow{2}{*}{48} \\  
    data cache {\scriptsize (KB)} \tiny $^{(\#)}$ & & & & & \\
    \hline
    L2 cache {\scriptsize (KB)} & 768 & 512 & \color{blue} 1536 &
    \color{blue} 1536 & \color{blue} 1536 \\
    \hline 
  \end{tabular}
  \caption{\label{tbl:spec}
    chip and memory specifications of recent NVIDIA GPUs
  }
\end{table}
Table~\ref{tbl:spec} presents chip and memory specification of NVIDIA
Kepler GPUs compared with Fermi GTX 580.
GTX Titan inherits most of important features of the Kepler
architecture such as new streaming multiprocessor SMX, increased
memory bandwidth and dynamic parallelism.
In addition, it supports the features provided only for Tesla or
Quadro GPUs such as large memory size and high performance in double
precision floating point calculation.

\begin{figure}[tbhp]
  \centering
  \begin{minipage}{0.4\linewidth}
    \begin{subfigure}[b]{0.9\linewidth}
      \includegraphics[width=1.0\linewidth]{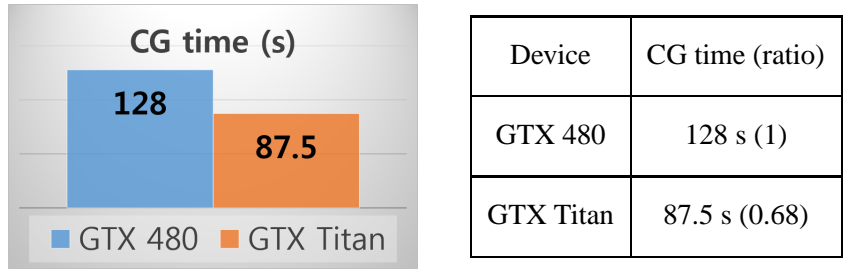}
    \end{subfigure}
  \end{minipage}
  \begin{minipage}{0.4\linewidth}
    \begin{subfigure}[b]{0.9\linewidth}
      \begin{tabular}{|c|c|}
        \hline
        \multirow{2}{*}{Device} &
        \multirow{2}{*}{CG time (ratio)}\\
        & \\
        \hline
        \multirow{2}{*}{GTX 480} & \multirow{2}{*}{128 s (1)} \\
        & \\
        \hline
        \multirow{2}{*}{GTX Titan} & \multirow{2}{*}{87.5 s (0.68)} \\
        & \\
        \hline
      \end{tabular}
    \end{subfigure}
  \end{minipage}
        
  \caption{CG performance by GTX 480 and GTX Titan. Here, CG time
    means the time (in the unit of second) which it takes to run the
    CG code 10 times in two GPUs.}
  \label{fig:cgNoOpt}
\end{figure}
Fig.~\ref{fig:cgNoOpt} shows the performance of conjugate gradient (CG)
solver by Fermi GTX 480 and Kepler GTX Titan without applying any
optimization to Kepler GPUs.
Although GTX Titan's performance are, in principle, much better than
that of GTX 480, our CG code is optimized only for Fermi GPUs and not
for Kepler GPUs.
Hence, it is necessary to tune the code such that it achieves the
highest performance for GTX Titan.

\begin{table}[tbhp]
  \centering
  \begin{tabular}{|c|>{\centering\arraybackslash}m{3cm}|>{\centering\arraybackslash}m{3cm}|}
    \hline
    & Fermi & Kepler \\
    \hline
    simultaneous blocks / SM(X) & 8 & 16 \\
    \hline
    warp schedulers / SM(X) & 2 & 4 \\
    \hline
    registers / thread & 63 & 255 \\
    \hline
    bandwidth (GB/s) & 192 (GTX 580) & 288 (GTX Titan) \\
    \hline
  \end{tabular}
  \caption{changed properties related with thread and block scheduling}
  \label{tbl:optSched}
\end{table}
There are several optimization schemes possible for GTX Titan.
Table~\ref{tbl:optSched} shows those changes in GPUs regarding thread
and block scheduling.
A SMX (Kepler) has 6 times more cores than SM (Fermi).
To deal with these cores, the SMX has twice number of blocks run
simultaneously and twice of warp schedulers than SM.
The number of registers per thread is also increased to 255, so that a
thread can store more variables to registers and reuse them quickly.
Therefore, we might obtain significantly better performance by simply
adjusting thread and block numbers.

The change of the memory bandwidth is also very important.
Unfortunately in general, the main bottle neck in GPUs are the
limitation in data transfer speed between GPU registers and memories.
The performance of a CUDA program is usually determined by the product
of CGMA (compute to global memory access) ratio and the amount of data
transfer per time \cite{Kirk:2010}.
Here, CGMA ratio is the number of floating point operations per single
data transfer.

Kepler architecture has new features to improve the memory usage as
follows.
\begin{itemize}
%
  \item 8 bytes shared memory bank mode is added.
    Fermi GPUs provide only 4 bytes (32 bits) mode.
    But Kepler GPUs provide 8 bytes (64 bits) mode, too.
    When this mode is turned on, one gets about twice the effective
    bandwidth for double precision floating point numbers.
    %
  \item It is possible to adjust the ratio of shared memory and L1
    cache to get the better performance with the total memory size
    fixed.
    Fermi GPUs only support 16 Kbytes (shared memory) + 48 Kbytes (L1
    cache) and 48 + 16 modes.
    Kepler GPUs can allocate 32 K to shared mem and 32 K to L1.
    %
  \item 48 KB Read-only data cache is added.
    The texture memory can be used as an additional read-only cache memory
    for Kepler GPUs.
    %
  \item Warp shuffle is introduced.
    By warp shuffle, data between threads in a warp can be exchanged
    without using shared memory.
    Thus we can reduce redundant use of the shared memory.
    Moreover, its latency is lower than shared memory access.
\end{itemize}

There is one more important new technology: direct parallelism.
If we use it, new threads can be spawned directly from GPU kernel, so
that one can reduce communications with CPU.
We are implementing the above new technologies to our GPU code.
%


%% file: gpuDir.tex
\section{GPU Direct}


Recently NVIDIA introduced an advanced version of GPU Direct, called
as GPU Direct RDMA (remote direct memory access).
GPU Direct is a technology by which one can improve communication
between GPUs, between a GPU and other network devices, and between a
GPU and storage devices.
There are three kinds of GPU Direct.
GPU Direct version 1 is designed for communication between GPU to
other network or storage devices.
GPU Direct version 2 provides peer-to-peer communication between GPUs
on the same PCIe bus.
They were already available in Fermi architecture.
The new one, GPU Direct RDMA extends this to infiniband network
communication between GPUs using RDMA.
Unfortunately, GPU Direct RDMA is only available for Tesla and Quadro
GPUs of Kepler architecture.
GTX Titan supports only GPU Direct v2.


\begin{figure}[tbhp]
  \centering
  \includegraphics[width=0.7\linewidth]{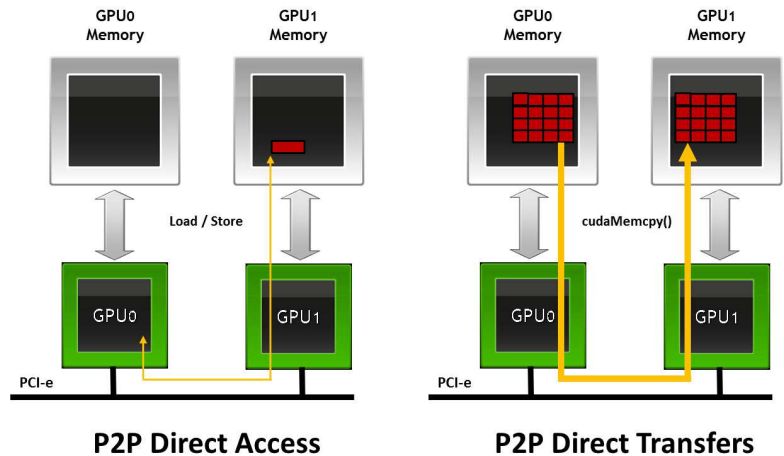}
  \caption{GPU Direct on the same PCIe bus}
  \label{fig:gpuDirect2}
\end{figure}
Fig.~\ref{fig:gpuDirect2} is the schematic diagram of GPU Direct v2,
by which a GPU can access the memory of another GPU on the same PCIe
bus or transfer data to the another without using the CPU memory.

A usual MPI + CUDA program assigns one GPU per one MPI process node.
For simplicity, consider a cluster node with 2 GPUs (GPU0 and GPU1) and we
run a MPI job of 2 processes.
The first MPI process (MPI0) is assigned with CPU0 and GPU0, and the second
one with CPU1 and GPU1.
Then we want to transfer some data stored in GPU0's memory to GPU1's
memory.
Without GPU Direct, we should follow a cumbersome procedure.
\begin{enumerate}
  \vspace{-0.3cm}
  \item First copy the data from GPU0's memory to CPU0's memory by using
    CUDA.
  \vspace{-0.3cm}
  \item Send the data in CPU0's memory to CPU1's memory through
    the memory of the infiniband network adapter by using MPI.
  \vspace{-0.3cm}
  \item Copy the data to GPU1's memory by using CUDA.
\end{enumerate}
Fig.~\ref{fig:withoutgd} shows the code doing this data transfer.
\begin{figure}[thbp]
 \centering
  \begin{lstlisting}
// GPU0 to CPU0
if( rank == 0 ) cudaMemcpyGtoC_host( a1, d_a, size );
// CPU0 to CPU1
if( rank == 0 ) MPI_Send( a1, N, MPI_FLOAT, 1, 0, MPI_COMM_WORLD );
else MPI_Recv( a1, N, MPI_FLOAT, 0, 0, MPI_COMM_WORLD, &status );
// CPU1 to GPU1
if( rank == 1 ) cudaMemcpyCtoG_host( d_a, a1, size );
  \end{lstlisting}
  \vspace{-0.4cm}
  \caption{data transfer without GPU Direct : GPU0 $\rightarrow$ CPU0
    $\rightarrow$ CPU1 $\rightarrow$ GPU1}
  \label{fig:withoutgd}
\end{figure}
Whereas, this 3-step data transfer can be reduced to a single step by
GPU Direct (Fig.~\ref{fig:withgd}).
With GPU Direct, the data in GPU0 are transferred to GPU1 at once by
a CUDA function: \texttt{cudaMemcpyPeer()}.
\begin{figure}[thbp]
  \centering
  \begin{lstlisting}
// GPU0 to GPU1
cudaMemcpyPeer( d_a1, 1, d_a0, 0, size );
  \end{lstlisting}
  \vspace{-0.4cm}
  \caption{data transfer with GPU Direct : GPU0 $\rightarrow$ GPU1}
  \label{fig:withgd}
\end{figure}

\begin{figure}[tbhp]
  \centering
  \begin{minipage}{0.5\linewidth}
    \begin{subfigure}[b]{0.9\linewidth}
      \includegraphics[width=1.0\linewidth]{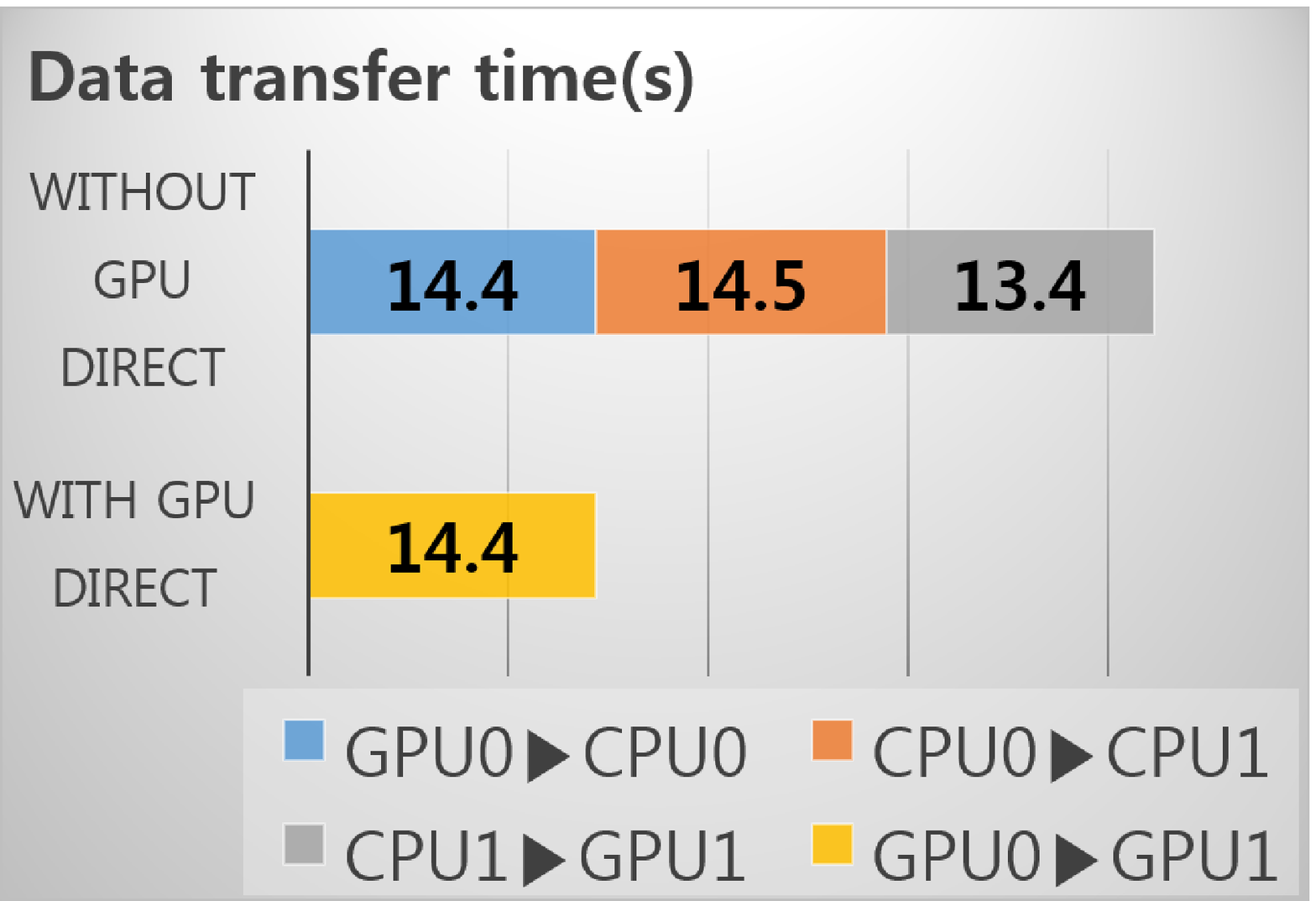}
    \end{subfigure}
  \end{minipage}
  \begin{minipage}{0.4\linewidth}
    \begin{subfigure}[b]{0.9\linewidth}
      \begin{tabular}{|c|c|}
        \hline
        \multirow{2}{*}{GPU Direct} &  \multirow{2}{*}{data transfer time
          (ratio)} \\
        & \\
        \hline
        \multirow{2}{*}{off} & \multirow{2}{*}{42.3 s (1)}\\
        & \\
        \hline
        \multirow{2}{*}{on} & \multirow{2}{*}{14.4 s ({\color{blue}0.34})}
        \\
        & \\
        \hline
      \end{tabular}
    \end{subfigure}
  \end{minipage}
  \caption{data transfer time for $10^7$ single precision numbers (40MB)
    from GPU0 to GPU1. Both GPUs are GTX Titan.}
  \label{fig:datatransfer}
\end{figure}
Fig.~\ref{fig:datatransfer} presents the transfer times for 40MB data
between 2 GTX Titan on the same PCIe bus with and without GPU Direct v2.
GPU Direct v2 reduces the transfer time down to 1/3.
Note that the data transfer between two GPUs by GPU Direct v2 takes
only the same amount of time as that between GPU and CPU by
\texttt{cudaMemcpy()}.
Hence, it costs only 1/3 of the full data transfer in the previous
method.


%% file: phi.tex
\section{Xeon Phi}


\begin{table}[h]
  \centering
  \begin{tabular}{|c|c|c|c|c|}
    \hline
    & \multicolumn{2}{c|}{Intel Xeon Phi} & \multicolumn{2}{c|}{NVIDIA GPU} \\    
    \hline
    & 7110X & 5110P & Tesla K20X & GTX Titan \\
    \hline
    \# of Cores & 61 & 60 & 2688 & 2688 \\
    \hline
    Core Clock (MHz) & 1333 & 1053 & 732 & 837 \\
    \hline
    SP TFLOPS & 2.44 & 2.02  & 3.95 & 4.5 \\
    \hline
    DP TFLOPS & \color{blue} 1.22 & \color{blue} 1.01 & \color{blue} 1.31 &
    \color{blue} 1.27 \\
    \hline
    Memory Size {\scriptsize (GB)} & \color{blue} 16 & \color{blue} 8 & 6 &
    6.1 \\
    \hline
    Mem. Bandwidth {\scriptsize (GB/s)} & \color{blue} 352 & \color{blue}
    320 & 250 & 288 \\
    \hline
    Price {\scriptsize (USD)} & 4130 & 2650 & 3800 & 1100 \\
    \hline
  \end{tabular}
  \caption{specification comparison between Intel Xeon Phi coprocessors and
    NVIDIA Kepler GPUs}
  \label{tbl:phiSpec}
\end{table}
Table~\ref{tbl:phiSpec} presents the specification of Xeon Phi
coprocessors compared with NVIDIA Kepler GPUs.
Although a Xeon Phi coprocessor has only 60 cores much less than those
(2688) in GPUs, each core is much faster than that of GPU.
A single core supports 512 bit SIMD (single instruction, multiple
data) operations, which allows simultaneous calculation of 8 double
precision floating point number operations.
Hence, a Xeon Phi coprocessor could, in principle, give similar
performance with GTX Titan theoretically.
More precisely, the (single precision) peak performance of GTX Titan
is
%
%
\begin{equation*}
  0.837 \textit{ (clock speed : GHz)} \times 2688 \textit{ (\# of cores)}
  \times 2 \textit{ (FMA)} = 4.50
  \textit{ (TFLOPS) .}
\end{equation*}
Here, FMA means fused multiply add instruction, which computes $a
\times x + y$ in one cycle.
Similarly, that of Xeon Phi 5110P is
%
%
\begin{equation*}
  1.053 \textit{ (clock speed : GHz)} \times 60 \textit{ (\# of cores)}
  \times 16 \textit{ (vectorization)} \times 2 \textit{ (FMA)} = 2.02
  \textit{ (TFLOPS) .}
\end{equation*}
In the case of double precision operations, Xeon Phi 5110P has about
the same performance ($\approx$1 TFlops) as GTX Titan ($\approx$1.3
TFlops).
However, in the case of Xeon Phi coprocessor, one should note that the
factor 16 boost comes from vectorization.
The vectorization means a parallelized calculation using SIMD instructions.
As already mentioned, Xeon Phi coprocessors support 512 bit SIMD
operations, so that 16 single precision calculations can be computed
simultaneously.
On the other hand, this means that the actual performance of Xeon Phi
system is highly dependent on the vectorization of the code.

However, it is \textbf{not always} possible to convert the code into a
vectorized one as specified by Xeon Phi systems.
The first difficulty is that one must program the code in the level
of the assembly language to control the array structure of the SIMD
registers \cite{Jeong:2012PoS}.
Unfortunately, the C level compiler cannot do this job automatically to
our satisfaction.
The second difficulty is that our QCD code is not, in general, designed
to fit it into the structure format of specific SIMD registers required 
by the vectorization.
This does not allow most of our code to be adapted to the
vectorization scheme.
Hence, in practice, this gain of 16 in vectorization is useless to us.
Therefore, in the end of day we find out that the real performance of
Xeon Phi systems is inferior to that of GTX Titan by a factor of about
10.
This result is quite discouraging.

\begin{figure}[tbhp]
  \centering
  \begin{minipage}{0.5\linewidth}
    \begin{subfigure}[b]{0.9\linewidth}
      \includegraphics[width=1.0\linewidth]{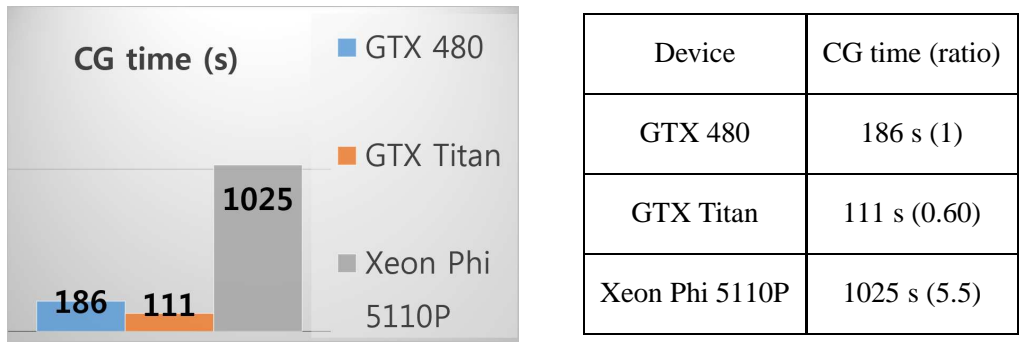}
    \end{subfigure}
  \end{minipage}
  \begin{minipage}{0.4\linewidth}
    \begin{subfigure}[b]{0.9\linewidth}
      \begin{tabular}{|c|c|}
        \hline
        \multirow{2}{*}{Device} & \multirow{2}{*}{CG time (ratio)} \\
        & \\
        \hline
        \multirow{2}{*}{GTX 480} & \multirow{2}{*}{186 s (1)} \\
        & \\
        \hline      
        \multirow{2}{*}{GTX Titan} & \multirow{2}{*}{111 s (0.60)} \\
        & \\
        \hline      
        \multirow{2}{*}{Xeon Phi 5110P} & \multirow{2}{*}{1025 s (5.5)} \\
        & \\
        \hline
      \end{tabular}
    \end{subfigure}
  \end{minipage}
        
  \caption{CG performance by GTX 480, GTX Titan, and Xeon Phi. Here, CG
    time means the time (in the unit of second) which it takes to run the
    CG code 10 times in a single GPU or a single Xeon Phi.}
  \label{fig:cgXeonPhi}
\end{figure}
Fig.~\ref{fig:cgXeonPhi} shows the performance of the CG solver by
Xeon Phi 5110P compared with the performance by GTX 480 and GTX Titan.
Even though the code is vectorized by Intel compiler, Xeon Phi 5110P solves
CG about 5 times and 10 times slower than GTX 480 and GTX Titan
respectively.


%% file: conclusion.tex
\section{Conclusion}


GTX Titan provides 1.15 GFLOPS per USD of double precision performance,
which is much better than 0.35 of Tesla K20X and 0.38 of Xeon Phi 5110P.
Besides the theoretical performance, there are many other improvements on
Kepler GPUs, such as direct parallelism, Hyper-Q and GPU Direct RDMA.
By applying GPU Direct v2 to two GPUs on the same PCIe bus, we
achieved about 3 times gain in data transfer.
We also investigated the Xeon Phi system.
However, the performance of Xeon Phi is so low (by a factor of 10)
that we do not recommend using Xeon Phi systems for the lattice QCD
simulation yet.


%% file: acknowledge.tex
\section{Acknowledgement}

The research of W.~Lee is supported by the Creative Research
Initiatives program (2013-003454) of the NRF grant funded by the
Korean government (MSIP).
This work was supported by SNU Undergraduate Research Program.
This work was supported by Seoul Science High School R\&E program.
W.~Lee acknowledges support from the KISTI supercomputing
center through the strategic support program [No.~KSC-2012-G3-08].
%